\documentclass[aps,prl,groupedaddress,showpacs,floatfix]{revtex4}

\usepackage{amsfonts}
\usepackage{graphicx}

\newtheorem{theorem}{Theorem}

\def\endprf{\hfill  {\vrule height6pt width6pt depth0pt}\medskip}
\newenvironment{proof}{\noindent {\bf Proof} }{\endprf\par}

\begin{document}


\title{Observability of glueball spectrum in QCD and the width of $\sigma$ resonance}


\author{Marco Frasca}
\email[]{marcofrasca@mclink.it}
\affiliation{Via Erasmo Gattamelata, 3 \\ 00176 Roma (Italy)}


\date{\today}

\begin{abstract}
We prove a theorem in QCD stating that in the limit of strong coupling,
$g\rightarrow\infty$, the observed spectrum of glueballs in QCD is the
same of a pure Yang-Mills theory, being mixing effects due to the
next-to-leading order. A full effective theory for QCD is obtained and the width of the
$\sigma$ resonance decay is straightforwardly computed. This appears as the lowest glueball state.
Vacuum gluon condensate is computed that consistently support studies on the identification of this meson as a glueball.
\end{abstract}

\pacs{11.15.-q, 11.15.Me}

\maketitle


The question of the light scalar meson spectrum is an open problem whose solution would imply a great improvement in our ability to manage low-energy QCD. This understanding is currently lacking and we have difficulties to comprehend the nature of a number of states that appear in laboratory. One of the main open questions is if glueballs can be seen and, if they have been seen, what observed states these are. This lack of knowledge has also implied serious difficulties into the identification of molecular or tetraquark states that some researchers assume representing the light scalar spectrum \cite{mai}. 
This view is challenged by some recent analysis by Mennessier, Narison, Ochs and Minkowski on the $\sigma$ reonance \cite{nar1,nar2} giving evidence for its glueball nature. More recently, Kaminski, Mennessier and Narison gave a stronger evidence from KK decay \cite{nar3}.
Indeed,  one of the most demanding problem is to know what is the structure of the lowest resonance seen so far, also presented as f0(600). Initially its very existence was strongly debated. Today, there are precise determinations from experimental data of its mass and width \cite{ygp,lcc} and is inserted into particle listing \cite{pdg}. Its observed decays are $\sigma\rightarrow\pi^+\pi^-$ and $\sigma\rightarrow\gamma\gamma$ being the former largely dominant. So, a great understanding of QCD would be achieved if we would be able to compute the mass and the width of this particle. The aim of this paper is to show how to obtain these results starting from QCD. Our main conclusion will be that this particle is indeed a glueball and the ground state of pure Yang-Mills theory. This result relies on an essential way to see how quarks and glueball state can mix in QCD. There is no serious proof about and is generally believed that, also in the low-energy limit, some mixing must happen. But we will prove a theorem stating that the observed glueball states should have almost exactly the properties of the states of a pure Yang-Mills theory and mixing is showed to be ineffective. 

The line of research started with the introduction of a gluon condensate \cite{svz} strongly support our conclusions. This means that there must be a deep connection between our results and those obtained by the aforementioned authors. We will show that a vacuum gluon condensate is indeed different from zero and has the expected value in agreement with recent analysis \cite{iof,ale}. This proves that all these studies are well founded and their conclusions correct.

So, we prove the following theorem that will give the concept of ordering needed for successive computations.

\begin{theorem}[Spectral Theorem]
\label{teo1}
At the leading order of a perturbation series of QCD for the coupling $g$ going to infinity
the observable spectrum of glueballs in QCD is the same of a pure Yang-Mills theory.
\end{theorem}

\begin{proof}
We take the following actions for the Yang-Mills field \cite{nair}
\begin{eqnarray}
    S_{YM}&=&-\int d^4x\left[\frac{1}{4}(\partial_\mu A^a_\nu-\partial_\nu A^a_\mu)
    (\partial^\mu A^{a\nu}-\partial^\nu A^{a\mu})
    +\frac{1}{2\alpha}(\partial\cdot A^a)^2
    +\partial^\mu\bar c^a\partial_\mu c^a\right] \\ \nonumber
    &-&\int d^4x\left[\frac{g}{2}f^{abc}(\partial_\mu A_\nu^a-\partial_\nu A_\mu^a)A^{b\mu}A^{c\nu}
	+\frac{g^2}{4}f^{abc}f^{ars}A^b_\mu A^c_\nu A^{r\mu}A^{s\nu}
	+gf^{abc}\partial_\mu\bar c^a A^{b\mu}c^c\right].
\end{eqnarray}
with $g$ the coupling constant and $\alpha$ fixing the gauge. Similarly, $f^{abc}$ are SU(N) structure constants
and $c^a$ is the ghost field. The action for the fermion fields is
\begin{equation}
   S_m = \int d^4x \sum_q \bar q(x)\left[i\gamma\cdot\left(\partial+ig\frac{\lambda^a}{2}A^a\right)-m_q\right]q(x)
\end{equation}
being $q$ the flavor index and $\lambda^a$ the group generators. So, the generating functional for the quantum field theory is
\begin{eqnarray}
   Z[\eta,\bar\eta,\epsilon,\bar\epsilon,j]&=&\int \prod_q[dq][d\bar q][dc][d\bar c][dA]
   e^{i(S_{YM}+S_m)} \times \\ \nonumber
   & &e^{i\int d^4x\prod_q[\bar q(x)\eta_q(x)+\bar\eta_q(x)q(x)]} \times \\ \nonumber
   & &e^{i\int d^4x[\bar c^a(x)\epsilon^a(x)+\bar\epsilon^a(x)c^a(x)]} \times \\ \nonumber
   & &e^{i\int d^4x j^a_\mu(x)A^{a\mu}(x)}.
\end{eqnarray}

In order to prove the theorem the choice of the gauge is not relevant and so we take one that makes Yang-Mills action simpler. We choose $\alpha = 1$ so that we can write \cite{nair}
\begin{eqnarray}
    S_{YM}&=&-\int d^4x\left[\frac{1}{2}\partial_\mu A^a_\nu\partial^\mu A^{a\nu}
    +\partial^\mu\bar c^a\partial_\mu c^a\right] \\ \nonumber
    &-&\int d^4x\left[gf^{abc}\partial_\mu A_\nu^aA^{b\mu}A^{c\nu}
	+\frac{g^2}{4}f^{abc}f^{ars}A^b_\mu A^c_\nu A^{r\mu}A^{s\nu}
	+gf^{abc}\partial_\mu\bar c^a A^{b\mu}c^c\right].
\end{eqnarray}
The reason for this choice will be clear in the following.

We need to make contact with the phenomenology at low energies. In this case we have a large coupling $g$ making ordinary weak perturbation theory not applicable. In order to uncover an asymptotic approximation in this regime, we consider the formal limit $g\rightarrow\infty$ that is dual to the weak perturbation limit $g\rightarrow 0$. So, for a generic field $\phi$ entering into the generating functional, we ask a formal solution series like
\begin{equation}
   Z[\eta,\bar\eta,\epsilon,\bar\epsilon,j] = \sum_{n=0}^\infty (\sqrt{N}g)^{-n}
   Z_n[\eta,\bar\eta,\epsilon,\bar\epsilon,j]
\end{equation}
but, apparently, this series cannot be straightforwardly obtained from the generating functional. An obvious way out to this impasse has been proposed in \cite{fra0,fra1,fra2,fra3}. We choose to rescale the time variable as $t\rightarrow \sqrt{N}gt$ being $Ng^2$ 't Hooft coupling. So, we rewrite the actions above as
\begin{eqnarray}
    S_{YM}&=&-\sqrt{N}g\int d^4x\left[\frac{1}{2}\partial_0 A^a_\nu\partial_0 A^{a\nu}
    +\partial_0\bar c^a\partial_0 c^a
    +\frac{1}{\sqrt{N}}f^{abc}\partial_0\bar c^a A^{b}_0c^c\right. \\ \nonumber
    &+&\left.\frac{1}{\sqrt{N}}f^{abc}\partial_0 A_\nu^aA^{b}_0A^{c\nu}    
    +\frac{1}{4N}f^{abc}f^{ars}A^b_\mu A^c_\nu A^{r\mu}A^{s\nu}\right] \\ \nonumber  
    &+&\frac{1}{\sqrt{N}}\int d^4x\left[f^{abc}\partial_iA_\nu^aA^{b}_iA^{c\nu}
    +f^{abc}\partial_i\bar c^a A^{b}_ic^c\right] \\ \nonumber
    &+&\frac{1}{\sqrt{N}g}\int d^4x\left[\frac{1}{2}\nabla A^a_\nu\cdot\nabla A^{a\nu}
    +\nabla\bar c^a\cdot\nabla c^a\right]   
\end{eqnarray}
and so, a perturbation expansion will produce a gradient expansion being gradient terms of higher order in the limit $g\rightarrow\infty$. Then, for the quark fields we have
\begin{eqnarray}
   S_m &=& \int d^4x \sum_q 
   \bar q(x)\left[\gamma_0\cdot\left(i\partial_0-\frac{\lambda^a}{2\sqrt{N}}A^a_0\right)
   +\frac{\lambda^a}{2\sqrt{N}}\gamma_i\cdot A^a_i\right]q(x) \\ \nonumber
   &-&\frac{1}{\sqrt{N}g}\int d^4x\sum_q\bar q(x)\left[{\bf \gamma}\cdot(-i\nabla)+m_q\right]q(x).
\end{eqnarray}

We can expand the generating functional as
\begin{eqnarray}
   Z[\eta,\bar\eta,\epsilon,\bar\epsilon,j]&=&{\cal N}\int \prod_q[dq][d\bar q][dc][d\bar c][dA]\times \\ \nonumber
    &&\exp\left\{-i\sqrt{N}g\int d^4x\left[\frac{1}{2}\partial_0 A^a_\nu\partial_0 A^{a\nu}
    +\partial_0\bar c^a\partial_0 c^a
    +\frac{1}{\sqrt{N}}f^{abc}\partial_0\bar c^a A^{b}_0c^c\right.\right. \\ \nonumber
    &+&\left.\left.\frac{1}{\sqrt{N}}f^{abc}\partial_0 A_\nu^aA^{b}_0A^{c\nu}    
    +\frac{1}{4N}f^{abc}f^{ars}A^b_\mu A^c_\nu A^{r\mu}A^{s\nu}\right]\right\}\times \\ \nonumber  
    &&\exp\left\{i\frac{1}{\sqrt{N}}\int d^4x\left[f^{abc}\partial_iA_\nu^aA^{b}_iA^{c\nu}
    +f^{abc}\partial_i\bar c^a A^{b}_ic^c\right]\right\}\times \\ \nonumber
    &&\exp\left\{\int d^4x \sum_q 
    \bar q(x)\left[\gamma_0\cdot\left(i\partial_0-\frac{\lambda^a}{2\sqrt{N}}A^a_0\right)
    +\frac{\lambda^a}{2\sqrt{N}}\gamma_i\cdot A^a_i\right]q(x)\right\}\times \\ \nonumber
    &&\exp\left\{\frac{i}{\sqrt{N}g}
    \int d^4x\prod_q\left[\bar q(x)\eta_q(x)+\bar\eta_q(x)q(x)\right]\right\} \times \\ \nonumber
    &&\exp\left\{\frac{i}{\sqrt{N}g}
    \int d^4x\left[\bar c^a(x)\epsilon^a(x)+\bar\epsilon^a(x)c^a(x)\right]\right\} \times \\ \nonumber
    &&\exp\left\{\frac{i}{\sqrt{N}g}\int d^4x j^a_\mu(x)A^{a\mu}(x)\right\} + O\left(\frac{1}{\sqrt{N}g}\right) 
\end{eqnarray}
where we have just kept terms of order $O\left(\sqrt{N}g\right)$ and $O(1)$. Now we do the key observation that the first term of the expansion, in the limit $g\rightarrow\infty$ is that of the pure Yang-Mills field that is $O\left(\sqrt{N}g\right)$ and effects due to quark fields are just $O(1)$. This implies that, in the given limit, the spectrum of the pure Yang-Mills field must be observed with negligible corrections due to the mixing with quark fields. The reason is that quark currents do not drive the Yang-Mills field at the leading order that behaves as it would be free.

Finally, corrections due to gradients just appear at order $O\left(\frac{1}{\sqrt{N}g}\right)$.

\end{proof}

Some considerations are in order. The limit $g\rightarrow\infty$ implies that the leading order term 
\begin{eqnarray}
   \int[dA][dc][d\bar c]&&
   \exp\left\{-i\sqrt{N}g\int d^4x\left[\frac{1}{2}\partial_0 A^{a}_\nu\partial_0 A^{a\nu}
    +\partial_0\bar c^{a}\partial_0 c^{a}
    \right.\right. \\ \nonumber
    &+&\frac{1}{\sqrt{N}}f^{abc}\partial_0\bar c^{a} A^{b}_0c^{c}
    +\frac{1}{\sqrt{N}}f^{abc}\partial_0 A_\nu^{a}A^{b}_0A^{c\nu} \\ \nonumber 
    &+&\left.\left.\frac{1}{4N}f^{abc}f^{ars}A^{b}_\mu A^{c}_\nu A^{r\mu}A^{s\nu}
    \right]\right\}
\end{eqnarray}
can be evaluated in the semiclassical approximation. So, if we know a proper classical solution of Yang-Mills equations we will be able to get both the spectrum and the propagators of the theory at this order.

In order to have an understanding of this part of the theory one should care about what are the classical solutions to be selected to work with in a quantum field theory. Classical Yang-Mills equations admit both integrable and chaotic solutions. Here we just put forward a conjecture, to be proved, that {\sl a quantum field theory can only exist with integrable classical solutions}. The reasons to believe this rely mostly on the computational opportunities that integrable solutions do provide. Classical Yang-Mills equations admit a lot of integrable solutions and this can be easily selected with the so called Smilga's choice and the mapping theorem \cite{fra4}. These solutions map Yang-Mills theory on a quartic massless scalar field theory. For our aims this is equivalent to
\begin{equation}
    A_\mu^a(x)=\eta^a_\mu\phi(x)
\end{equation}
with $\phi(x)$ a solution of the equation $\partial_t^2\phi+Ng^2\phi^3=0$ and e.g.
\begin{equation}
    \eta^a_\mu=((0,1,0,0),(0,0,1,0),(0,0,0,1))
\end{equation}
for SU(2) and
\begin{equation}
    \eta^a_\mu=((0,0,0,0),(0,1,0,0),(0,1,0,0),(0,0,1,0),(0,1,1,0),(0,0,1,0),(0,0,0,1),(0,0,0,1))
\end{equation}
for SU(3) but the choices are a very large number and increase choosing a larger group. 
Then we can apply a Lorentz transformation and we will get a mapping solution between a scalar field and Yang-Mills theory.
For our aims 
it is enough
to consider the solution of the scalar field equation removing the gradient part. Then, starting with integrable solutions we now compute the Green function at the leading order. We have
\begin{eqnarray}
    \left.\frac{1}{Z}\frac{\delta Z}{\delta j^a_\mu(x)\delta j^b_\nu(y)}
    \right|_{\eta,\bar\eta,\epsilon,\bar\epsilon,j=0}&=&
    \frac{1}{Z[0]}\int[dA][dc][d\bar c]A^a_\mu(x)A^b_\nu(y) \\ \nonumber
    &&\exp\left\{-i\sqrt{N}g\int d^4x\left[\frac{1}{2}\partial_0 A^{a}_\nu\partial_0 A^{a\nu}
    +\partial_0\bar c^{a}\partial_0 c^{a}\right.\right. \\ \nonumber
    &+&\frac{1}{\sqrt{N}}f^{abc}\partial_0\bar c^{a} A^{b}_0c^{c}
    +\frac{1}{\sqrt{N}}f^{abc}\partial_0 A_\nu^{a}A^{b}_0A^{c\nu} \\ \nonumber 
    &+&\left.\left.\frac{1}{4N}f^{abc}f^{ars}A^{b}_\mu A^{c}_\nu A^{r\mu}A^{s\nu}\right]\right\}
    +O\left(\frac{1}{\sqrt{N}g}\right).
\end{eqnarray}
We can apply to this the above mapping theorem using the aforementioned integrable solutions giving
\begin{eqnarray}
    \left.\frac{1}{Z}\frac{\delta Z}{\delta j^a_\mu(x)\delta j^b_\nu(y)}
    \right|_{j=0}&=&\frac{1}{Z[0]}\int[dA][dc][d\bar c]
    \eta^a_\mu\eta^b_\nu\phi(x)\phi(y)\times \\ \nonumber
    &&\exp\left\{i\sqrt{N}g(N^2-1)\int d^4x\left[\frac{1}{2}\partial_0\phi\partial_0\phi
    +\partial_0\bar c^{a(0)}\partial_0 c^{a(0)}-\frac{1}{4}\phi^4\right]\right\} \\ \nonumber
    &+&O\left(\frac{1}{\sqrt{N}g}\right).
\end{eqnarray}
with really noteworthy simplifications. It is interesting to see how the ghost field becomes decoupled from the Yang-Mills field and so it is that of a free particle. These results were already seen in \cite{fra4}. A Green function can be considered also in the limit $g\rightarrow\infty$ as \cite{fra2,fra4}
\begin{equation}
    \partial_t^2 G(t)+Ng^2G(t)=\Lambda^2\delta(t)
\end{equation}
being $\Lambda$ an arbitrary constant having the dimension of energy. This is an integration constant of the theory and should be experimentally determined but an higher order theory should be able to compute it. The solution is easily written down as
\begin{equation}
\label{eq:G}
    G(t) = \theta(t)\Lambda\left(\frac{2}{Ng^2}\right)^{1\over 4}
    {\rm sn}\left[\left(\frac{Ng^2}{2}\right)^{1\over 4}\Lambda t\right]
\end{equation}
and the glueball spectrum is given through the Fourier series of the Jacobi sn function by
\begin{equation}
    m_n=(2n+1)\frac{\pi}{2K(i)}\sqrt{\sigma}
\end{equation}
being $K(i)\approx 1.311028777$ an elliptic integral and
we have identified $\sigma=\left(\frac{Ng^2}{2}\right)^{1\over 2}\Lambda^2$ as the string tension of the theory. This implies that, for the same value of $\Lambda$, $\sigma_{SU(2)}=\sqrt{2/3}\sigma_{SU(3)}$. This has been verified some years ago with lattice computations \cite{tep1}.

This spectrum, as is and together with the theorem above, makes a strong prediction. We can exploit this if we identify a set of pure numbers as
\begin{equation}
    a_n=\frac{m_n}{\sqrt{\sigma}}=(2n+1)\frac{\pi}{2K(i)}
\end{equation}
the first ones being given in Tab. \ref{tab:0++} compared to lattice computations \cite{tep2}. 
\begin{table}[tbp]
\begin{tabular}{|c|c|c|c|} \hline\hline
Excitation & Lattice & Theoretical & Error \\ \hline
$\sigma$   & -       & 1.198140235 & - \\ \hline 
0$^{++}$   & 3.55(7) & 3.594420705 & 1\% \\ \hline
0$^{++*}$  & 5.69(10)& 5.990701175 & 5\% \\ \hline\hline
\end{tabular}
\caption{\label{tab:0++} Comparison for the 0$^{++}$ glueball spectrum for SU(3).}
\end{table}
The most striking result is the existence of a lower ground state that can be identified with the observed $\sigma$ resonance at about 500 MeV \cite{pdg}. As said above, the very nature of this resonance is hotly debated yet \cite{kle} and is not seen in any lattice computation \cite{mcn}. From the above theorem in QCD together with our solution of Yang-Mills quantum field theory \cite{fra4} we can conlude immediately that this lowest state in the spectrum is a glueball. Further experimental analyses is needed to clarify this point. Anyhow, on the basis of the theorem we have just proved, the full glueball spectrum must be seen experimentally.

We want to pursue such an identification further. $\sigma$ resonance is known to have a quite large width. So, we estimate it evaluating the next-to-leading order correction to the solution presented here. Using again the mapping theorem \cite{fra4} and the integrable classical solutions of Yang-Mills theory, we can write down a full generating functional for an effective theory of infrared QCD. We will have
\begin{eqnarray}
   Z[\eta,\bar\eta,j_\phi] = \int[d\phi]\prod_q[dq][d\bar q]
    &&\exp\left\{i\sqrt{N}g(N^2-1)
    \int d^4x\left[\frac{1}{2}\partial_0\phi\partial_0\phi-\frac{1}{4}\phi^4\right]\right\} \times \\ \nonumber    
    &&\exp\left\{-i\frac{N^2-1}{\sqrt{N}g}\int d^4x\frac{1}{2}\nabla\phi\cdot\nabla\phi\right\}\times \\ \nonumber   
    &&\exp\left\{-i\frac{1}{\sqrt{N}g}\int d^4x\sum_q\bar q(x)\left[{\bf \gamma}\cdot(-i\nabla)+m_q\right]q(x)\right\}
    \times \\ \nonumber
    &&\exp\left\{i\int d^4x\sum_q 
    \bar q(x)\left[\gamma_0i\partial_0+\frac{\lambda^a}{2\sqrt{N}}
    \gamma_i\eta_i^a\phi(x)\right]q(x)\right\} \times \\ \nonumber
    &&\exp\left\{\frac{i}{\sqrt{N}g}\int d^4x\sum_q[\bar q(x)\eta_q(x)+\bar\eta_q(x)q(x)]\right\} \times \\ \nonumber
    &&\exp\left\{\frac{i(N^2-1)}{\sqrt{N}g}\int d^4x j_\phi\phi\right\}.    
\end{eqnarray}
having introduced $(N^2-1)j_\phi=j^a_\mu\eta^{a\mu}$. Then, we consider the following approximation that holds just in the infrared limit \cite{fra5,fra6}, this is indeed the leading order of a small time expansion,
\begin{equation}
    \phi(x)\approx \frac{1}{\sqrt{N}g}\int d^4y\Delta(x-y)j_\phi(y)
\end{equation}
where the multiplicative factor is there to account for rescaling in time and \cite{fra2}
\begin{equation}
    \Delta(x-y)=\delta^3(x-y)\left[G(t_2-t_1)+G(t_1-t_2)\right]
\end{equation}
is the Feynman propagator (see eq.(\ref{eq:G})). Such a small time approximation turns the functional for the scalar field into a Gaussian form \cite{fra2}. In order to show this, we compute the generating functional taking
\begin{equation}
    \phi = \phi_0+\frac{1}{\sqrt{N}g}\phi_1+O\left(\frac{1}{Ng^2}\right)     
\end{equation}
being
\begin{equation}
    \phi_0=\frac{1}{\sqrt{N}g}\int d^4y\Delta(x-y)j_\phi(y)
\end{equation}
where use is made of the aforementioned small time expansion. Working in this approximation, we can neglect terms of order higher than second as these imply higher powers of time variable. So, removing the rescaling in time, we are left with the following Gaussian approximation
\begin{eqnarray}
    Z[\eta,\bar\eta,j_\phi]&\approx&
    \exp\left\{\frac{i}{2}(N^2-1)\int d^4xd^4y j_\phi(x)\Delta(x-y)j_\phi(y)\right\}\times \\ \nonumber
    &&\int\prod_q[dq][d\bar q]
    \exp\left\{i\int d^4x\sum_q 
    \bar q(x)\left[i\gamma_0\partial_0+g\frac{\lambda^a}{2}
    \gamma_i\eta_i^a\int d^4y \Delta(x-y)j_\phi(y)\right]q(x)\right\}\times \\ \nonumber
    &&\exp\left\{i\int d^4x\sum_q[\bar q(x)\eta_q(x)+\bar\eta_q(x)q(x)]\right\}.
\end{eqnarray}
The remaining integral is Gaussian and we will be able to compute it when we know the quark propagator obtained by
solving the equation
\begin{equation}
    \left[i\gamma_0\partial_0+g\frac{\lambda^a}{2}\gamma_i\eta_i^a\int d^4y \Delta(x-y)j_\phi(y)\right]S[j_\phi,x]=
    \delta^4(x)
\end{equation}
to give us finally the following generating functional
\begin{eqnarray}
    Z[\eta,\bar\eta,j_\phi]&\approx&
    \exp\left\{\frac{i}{2}(N^2-1)\int d^4xd^4y j_\phi(x)\Delta(x-y)j_\phi(y)\right\}\times \\ \nonumber
    &&\exp\left\{i\int d^4xd^4y\sum_q\bar\eta_q(x)S[j_\phi,x-y]\eta_q(y)\right\}
\end{eqnarray}
and so our final aim will be to compute this functional of $j_\phi$. Indeed, this can be straightforwardly obtained as
\begin{equation}
    S[j_\phi,x-y]=\theta(t_x-t_y)\delta^3(x-y)\exp\left\{ig\frac{\lambda^a}{2}
    \gamma_0\gamma_i\eta_i^a\int_{t_x}^{t_y} dt'
    \int d^4x_1 \Delta(t'-t_y-t_{x_1},x-y-x_1)j_\phi(x_1)\right\}.
\end{equation}
At this stage we do the Nambu-Jona-Lasinio approximation on the gluon propagator \cite{fra7,fra8}
\begin{equation}
    \Delta(x-y)\approx\frac{3.76}{\sigma}\delta^4(x-y)
\end{equation}
being $\sigma$ the string tension. So, one has
\begin{equation}
    S[j_\phi,x-y]=\theta(t_x-t_y)\delta^3(x-y)\exp\left\{iG_\phi\frac{\lambda^a}{2}
    \gamma_0\gamma_i\eta_i^a\int_{t_x}^{t_y} dt'j_\phi(t'-t_y,x-y)\right\}.
\end{equation}
being the coupling $G_\phi=G_{NJL}/\sqrt{4\pi\alpha_s}$ and $G_{NJL}=3.76\frac{g^2}{\sigma}$ the Nambu-Jona-Lasinio coupling.

From this we can evaluate the vertex at a tree level as
\begin{eqnarray}
    &&\left.i\frac{\delta}{\delta j_\phi(x)}\frac{\delta}{i\delta\bar\eta_u(x)}\frac{i\delta}{\delta\eta_d(x)}
    \frac{\delta}{i\delta\bar\eta_d(x)}
    \frac{i\delta}{\delta\eta_u(x)}Z[\eta,\bar\eta,j_\phi]\right|_{j_\phi,\bar\eta,\eta=0}
    = \\ \nonumber
    &-&iG_\phi\frac{\lambda^a}{2}\gamma_0\gamma_i\eta_i^a \times \\ \nonumber
    &&\left[\theta(t_2-t_3)\theta(t_2-t_3-t_1)
  \delta^3(x_2-x_3)\delta^3(x_1)
  -\theta(t_4-t_5)\theta(t_4-t_5-t_1)\delta^3(x_4-x_5)\delta^3(x_1)\right]
\end{eqnarray}
corresponding to the process $\sigma\rightarrow\pi^+\pi^-$. In order to evaluate this vertex we use the following expression for the Heaviside function
\begin{equation}
   \theta(t)=-\frac{1}{2\pi i}\int_{-\infty}^{+\infty}dE\frac{1}{E+i0}e^{-iEt}.
\end{equation}
We have to go from this n-point function to the probability amplitude introducing pion fields. We do this with the following rule: we remove the contributions from the Heaviside functions with the product $m_\sigma f_\pi^2$, assuming $\sigma$ at rest and $m_\sigma$ its mass and $f_\pi$ the pion decay constant being about 93 MeV.  This corresponds to LSZ reduction. So, when we take the trace of the square of the amplitude, we will be left with
\begin{equation}
   |{\cal M}_{if}|^2=N(N^2-1)DG^2_\phi m_\sigma^2f_\pi^4
\end{equation}
that we specialize to $N=3$ and $D=4$. This will give the rate
\begin{equation}
   \Gamma_\sigma = \frac{6}{\pi}\frac{G^2_{NJL}}{4\pi\alpha_s} m_\sigma f_\pi^4\sqrt{1-\frac{4m_\pi^2}{m_\sigma^2}}
\end{equation}
that is in agreement with recent derivations from experiments \cite{ygp} being $\Gamma_\sigma/2=255\pm 10\ MeV$ for a mass $m_\sigma=484\pm 17\ MeV$ when $\alpha_s\approx 1.4$. A similar conclusion can be drawn also with respect to the derivation given in \cite{lcc}, presenting a mass $m_\sigma=441^{+16}_{-8}\ MeV$ and $\Gamma_\sigma/2=279^{+9}_{-12.5}\ MeV$, for $\alpha_s\approx 1.8$ showing that both results are consistent each other with respect to physical values of the strong coupling constant. We have kept the string tension fixed in both the computations at $(440\ MeV)^2$ but this choice may be too tightening with respect to the mass of the particle.

Our next step is to show how a gluon condensate emerges from our strong coupling computations. This step is an essential one as several authors used this concept to show that $\sigma$ meson is indeed a glueball with a mass in agreement with the one we obtained above. Indeed, we can show that there is a gluon condensate and its value is in close agrement with recent estimations \cite{iof,ale} and, in any case, very close to the value estimated by Shifman, Vainshtein and Zakharov \cite{svz}. So, we take SU(3) for the gauge group and apply the mapping theorem for Yang-Mills theory and a quartic scalar field theory \cite{fra4} giving
\begin{equation}
   \langle G\cdot G\rangle 
   = 16\langle (\partial\phi)^2\rangle-96\pi\alpha_s\langle\phi^4\rangle.
\end{equation}
This equation implies that we have to compute the following integrals
\begin{eqnarray}
   \langle (\partial\phi)^2\rangle&=&\int\frac{d^4p}{(2\pi)^4}p^2\Delta(p) \\ \nonumber
   \langle\phi^4\rangle&=&-3\left[\int\frac{d^4p}{(2\pi)^4}\Delta(p)\right]^2
\end{eqnarray}
that are not defined unless we introduce a cut-off. This is normally done in this kind of computations and, as seen above, QCD has a natural cut-off by its own that makes these integrals meaningful. So, for the gluon propagator, we introduce also a dependence on the space momentum after resumming the gradient part of the action that can always be done and introducing a mass $M$ that, for comparison with existing data, we take to be the mass of the charmed quark 1.275 GeV \cite{iof,ale}. Fourier transform of the gluon propagator is
\begin{equation}
\label{eq:prop}
    \Delta(p)=\sum_{n=0}^\infty\frac{B_n}{p^2-m_n^2+i\epsilon}
\end{equation}
being
\begin{equation}
    B_n=(2n+1)\frac{\pi^2}{K^2(i)}\frac{(-1)^{n+1}e^{-(n+\frac{1}{2})\pi}}{1+e^{-(2n+1)\pi}}.
\end{equation}
So, we finally get
\begin{equation}
\langle\frac{\alpha_s}{\pi}G\cdot G\rangle=
\alpha_s\frac{M^4}{2\pi^3}\left(0.2625+0.3304\frac{\alpha_s}{8\pi}\right)
\end{equation}
yielding our final result
\begin{equation}
\langle\frac{\alpha_s}{\pi}G\cdot G\rangle=(0.0039\pm 0.0003)GeV^4
\end{equation}
having taken $\alpha_s(m_\tau)=0.34\pm 0.03$. This is fully consistent with the bound
\begin{equation}
\langle\frac{\alpha_s}{\pi}G\cdot G\rangle < 0.008 GeV^4
\end{equation}
and values currently known \cite{iof,ale}. But this is an evidence obtained directly from QCD of the existence of the gluon condensate that gives a firm ground to all the conclusions drawn assuming its existence. Let us point out that some recent analysis point toward an higher value for the gluon condensate \cite{nar4} and this question is currently matter of debate.

We have presented a proof of the observability of the pure Yang-Mills spectrum in current experiments. The theorem so proved permitted to obtain an effective theory to compute the width of the $\sigma$ resonance that in this way appears clearly as the ground state of Yang-Mills theory. The obtained width is in close agreement with experimental data for proper values of the strong couplig constant. Finally, we showed the existence of a gluon condensate that firmly estabilishes the identification of $\sigma$ already obtained in this way.



\begin{thebibliography}{99}
\bibitem{mai} G. 't Hooft, G. Isidori, L. Maiani, A.D. Polosa, V. Riquer, Phys. Lett. {\bf B662}, 424 (2008).
\bibitem{nar1} G. Mennessier, S. Narison, W. Ochs, Phys. Lett. {\bf B665}, 205 (2008).
\bibitem{nar2} G. Mennessier, P. Minkowski, S. Narison, W. Ochs, arXiv:0707.4511 [hep-ph].
\bibitem{nar3} R. Kaminski, G. Mennessier, S. Narison, arXiv:0904.2555 [hep-ph].
\bibitem{ygp} F. J. Yndurain, R. Garcia-Martin, J. R. Pelaez, Phys. Rev. D 76, 074034 (2007).
\bibitem{lcc} I. Caprini, G. Colangelo, Leutwyler H., Phys. Rev. Lett. 96, 132001 (2006).
\bibitem{pdg} C. Amsler et al. (Particle Data Group), Phys. Lett. {\bf B667}, 1 (2008).
\bibitem{svz} M. A. Shifman, A. I. Vainshtein, V. I. Zakharov, Nucl. Phys. B147, 385, 448 and 519 (1979);
\bibitem{iof} B. L. Ioffe, Prog. Part. Nucl. Phys. {\bf 56}, 232 (2006).
\bibitem{ale} M. Davier, A. Hoecker, Zhiqing Zhang, Nucl. Phys. B: Proc. Suppl. {\bf 169}, 22 (2007).
\bibitem{nair} V. P. Nair, {\sl Quantum Field Theory}, (Springer, New York, 2005).
\bibitem{fra0} M. Frasca, Phys. Rev. A {\bf 58}, 3439 (1998). 
\bibitem{fra1} M. Frasca, Int. J. Mod. Phys. D {\bf 15}, 1373 (2006).
\bibitem{fra2} M. Frasca, Phys. Rev. D {\bf 73}, 027701 (2006); Erratum-ibid. D {\bf 73} 049902 (2006).
\bibitem{fra3} M. Frasca, Int. J. Mod. Phys. A {\bf 22} 1727 (2007).
\bibitem{fra4} M. Frasca, Phys. Lett. {\bf B670}, 73 (2008).
\bibitem{tep1} J. B. Kogut, D. K. Sinclair, M. Teper, Phys. Rev. D {\bf 44}, 2869 (1991).
\bibitem{tep2} B. Lucini, M. Teper, U. Wenger, JHEP 06, 012 (2004).
\bibitem{kle} E. Klempt, A. Zaitsev, Phys. Rep. {\bf 454}, 1 (2007).
\bibitem{mcn} C. McNeile, Nucl. Phys. B: Proc. Suppl. {\bf 186}, 264 (2009).
\bibitem{fra5} M. Frasca, Mod. Phys. Lett. A 22, (2007) 1293.
\bibitem{fra6} M. Frasca, Int. J. Mod. Phys. A 23 (2008) 299.
\bibitem{fra7} M. Frasca, arXiv:0803.0319 [hep-th], to appear in International Journal of Modern Physics E.
\bibitem{fra8} M. Frasca, Nucl. Phys. B: Proc. Suppl. {\bf 186}, 260 (2009).
\bibitem{nar4} S. Narison, Phys. Lett. {\bf B673}, 30 (2009).
\end{thebibliography}
\end{document}